\documentclass[10pt]{article}%
\usepackage{latexsym,delarray,multicol}
\usepackage{epsfig}
\usepackage{amsfonts}
\usepackage{amsmath}
\usepackage{amssymb}
\usepackage{revsymb}
\usepackage{graphicx}
\usepackage{url}%
\newcommand{\nc}{\newcommand}
\nc{\nn}{\nonumber} \nc{\ep}{\varepsilon} \nc{\la}{\lambda}
\nc{\wht}{\widehat} \nc{\ov}{\overline} \nc{\ds}{\displaystyle}
\nc{\ts}{\textstyle}
\nc{\kro}{\left(}\nc{\kvo}{\left[}\nc{\fio}{\left\{}
\nc{\krz}{\right)}\nc{\kvz}{\right]}\nc{\fiz}{\right\}}

\begin{document}

\title{On decoding algorithms for polar codes\vspace{-0.1in}}
\author{Ilya Dumer \thanks{\vspace{-0.04in}I. Dumer is with ECE Department, University
of California, Riverside, USA; email: dumer@ee.ucr.edu }}
\date{}
\maketitle

\begin{abstract}
\noindent We survey the known list decoding algorithms for polar codes and
compare their complexity.\vspace{0.02in}

\noindent\textbf{Index terms: }Polar codes; Reed-Muller codes; successive
cancellation decoding.\vspace{0.05in}

\end{abstract}

\thispagestyle{empty}

\vspace{-0.05in}

\section{\textbf{A brief survey of recursive decoding algorithms }\label{sub:1}}

Successive cancellation (SC) decoding was considered in \cite{dum1999} for
general Reed-Muller codes $RM(r,m)$ of order $r$ and dimension $m$. It was
also proposed in \cite{dum1999} to set to zeros those information bits that
are the least protected in SC decoding. Simulation results of \cite{dum1999}
show that the resulting subcodes with frozen bits significantly outperform the
original codes $RM(r,m).$ Subsequent papers \cite{dum2000}-\cite{dum2002}
extend SC technique to the list decoding of RM codes and their bit-frozen
subcodes. Simulation results of these papers show that the optimal selection
of the frozen bits in RM codes brings SC list decoding close to the maximum
likelihood decoding for the code lengths $n\leq512$ and list sizes $L\leq
64.$\vspace{0.03in}\vspace{0.02in}

A breakthrough in this area was achieved by E. Arikan \cite{ari}, who proved
that the bit-frozen subcodes of the full code $RM(m,m)$ - now well known as
polar codes - achieve the channel capacity of a symmetric memoryless channel
as $m\rightarrow\infty$. Paper \cite{ari} also employs a novel analytical
technique and reveals new properties of probabilistic recursive processing,
such as bit polarization. We also note that the specific choice of the maximal
order $r=m$ is immaterial in this case since the results of \cite{ari} hold
for the optimized bit-frozen subcodes of any code $RM(r,m)$ of rate $R\to1,$
in particular if $r/m>1/2$ for $m\rightarrow\infty$.\vspace{0.03in}%
\vspace{0.02in}

SC list decoding was later considered in \cite{vardy1}. This paper cites a
similar algorithm of \cite{dum2006} but relates the algorithm of
\cite{dum2006} to RM codes only. This is incorrect. All papers \cite{dum2000}%
-\cite{dum2006} address list decoding of the optimized bit-frozen subcodes,
and all emphasize large improvements that these subcodes achieve over the
original RM codes. Paper \cite{vardy1} also incorrectly asserts that the
recursive processing of $L$ codewords in \cite{dum2006} may require $n^{2}\log
n$ operations per one codeword, as opposed to only $n\log n$ operations for
$L=1.$ In fact, SC list decoding of \cite{dum2000}-\cite{dum2006} yields
complexity of $n\log n$ per one codeword and the overall complexity has the order of $Ln\log
n$ for both RM codes and their subcodes.\vspace{0.03in}\vspace{0.02in}

In summary, papers \cite{dum2000}-\cite{dum2006} and \cite{vardy1} use a
similar decoding algorithm and apply it to the same class of the bit-frozen
subcodes of RM codes. On the other hand, we also note that the design of polar
codes in \cite{vardy1} complements the earlier constructions of polar codes by
using the precoded information blocks and the fast analytical technique of
\cite{vardy2}, which gives the  output bit error rates without any simulation.\vspace{0.05in}

Below, we discuss recursive design and decoding of polar codes in more detail.

\section{\textbf{Recursive design of RM and polar codes}\label{sub:2}}

We first describe some recursive properties of RM\ or polar codes similarly to
papers \cite{dum2000,dum2006}. Consider boolean polynomials $f(x)$ of degree
$r$ or less in $m$ binary variables $x_{1},\ldots,x_{m}$, where $r\leq
m$.\ Vectors $x=(x_{1},...,x_{m})$ will mark the positions of our code. We
also use short notation $x_{i\,|\,j}=(x_{i},...,x_{j})$ for a punctured vector
$x$, where $i\leq j.$ \ Each map $f(x):\mathbb{F}_{2}^{m}\rightarrow
\mathbb{F}_{2}$ generates a codeword $\mathbf{c=c}(f)$ of a code $RM(r,m).$
\ Below, we take any sequence $(i_{1},...,i_{m})\in\mathbb{F}_{2}^{m}$ and
describe the recursive decomposition%
\begin{equation}%
\begin{tabular}
[c]{l}%
$f(x)=f_{0}(\mathbf{x}_{2\,|\,m})+x_{1}f_{1}(\mathbf{x}_{2\,|\,m}%
)=...\smallskip$\\
$=\sum_{i_{1},...,i_{\ell}}x_{1}^{i_{1}}\cdot...\cdot x_{\ell}^{i_{\ell}%
}\,\,f_{i_{1},...,i_{\ell}}(\mathbf{x}_{\ell+1\,|\,m})\smallskip$\\
$=...=\sum_{i_{1},...,i_{m}}f_{i_{1},...,i_{m}}\;x_{1}^{i_{1}}\cdot...\cdot
x_{m}^{i_{m}}\smallskip$%
\end{tabular}
\ \ \ \ \ \ \ \ \ \label{poly1}%
\end{equation}
Two polynomials $f_{0}(\mathbf{x}_{2\,|\,m})$ and $x_{1}f_{1}(\mathbf{x}%
_{2\,|\,m})$ derived in the first step generate codewords $\left(  \mathbf{c}%
_{0},\mathbf{c}_{0}\right)  $ and $\left(  \mathbf{0},\mathbf{c}_{1}\right)  $
\ in $\mathbb{F}_{2}^{m}.$ For each $i_{1}=0,1,$ the codeword $\mathbf{c}_{i_{1}}$ belongs to
the code $RM(r-i_{1},m-1)$. This yields the Plotkin
construction $\mathbf{c=c}_{0}\mathbf{,c}_{0}\mathbf{+c}_{1}$ of code
$RM(r,m)$. \ Similarly, any step $\ell=2,...,m$ decomposes the polynomial
$f(x)$ with respect to various monomials $x_{1}^{i_{1}}\cdot...\cdot x_{\ell
}^{i_{\ell}}$. We define these monomials using the binary strings $\xi_{1\,|\,\ell}=i_{1},...,i_{\ell}$, which we call binary
\textit{paths} of length $\ell $. For each path
$\xi_{1\,|\,\ell},$ the remaining part $f_{i_{1},...,i_{\ell}}(\mathbf{x}%
_{\ell+1\,|\,m})$ of decomposition (\ref{poly1}) defines some codeword $\mathbf{c}_{i_{1},...,i_{\ell}}$ of
length $2^{m-\ell}.$ Finally, any \textit{full path} $\xi=(i_{1},...,i_{m})$
of step $m$ defines the single monomial
\[
x^{\xi}\equiv x_{1}^{i_{1}}\,\cdot...\cdot\,x_{m}^{i_{m}}%
\]
which has  some coefficient $f\left(\xi\right)  =f_{i_{1},...,i_{m}}=0,1.$
Note that any path $\xi$ that ends with an information bit $f\left(
\xi\right)  =1$ gives some vector $\mathbf{c\mathbf{(\xi)}}\equiv
\mathbf{c}(x^{\xi})$ of length $n$ and weight $2^{m-w(\xi)},$ where
$w(\xi)$ is the Hamming weight of the string $\xi.$ If $f\left(\xi\right)
=0,$ then $\mathbf{c}(\xi)=0.$ RM codes $RM(r,m)$ include only $k(r,m)$ paths
of weight $w(\xi)\leq r$, where $k(r,m)$ is the dimension of the code $RM(r,m)$. 

Decomposition (\ref{poly1}) is also shown in Fig. 1 and 2. Here the full code
$RM(4,4)$ is depicted in Fig. 1. Each decomposition step $\ell=1,...,4$ is
marked by the splitting monomial $x_{\ell}^{i_{\ell}}.$ For example, path
$\xi=0110$ gives the coefficient $f_{0110}$ \ associated with the monomial
$x^{\xi}\equiv x_{2}x_{3}$.

Fig. 2 depicts code $RM(2,5).$ Here we only include all paths
$\mathbf{\mathbf{\mathbf{\xi}}}$ of weight $w(\xi)\leq2.$ Note that any two
paths $\xi_{1\,|\,\ell}$ entering some node have the same weight $w$ and
generate the same code $RM(r-w,m-\ell)$ on their extensions $\xi_{1\,|\,\ell
},i_{\ell+1},...,i_{m}$. For example, path $\xi=01100$ proceeds from $RM(2,5)$
to the single bit $RM(0,0)$ via nodes $RM(2,4),$ $RM(1,3),$ $RM(0,2),$ and
$RM(0,1).$

\begin{figure}[ptb]
\vspace{-0.1in}
\includegraphics[width=0.48\textwidth]{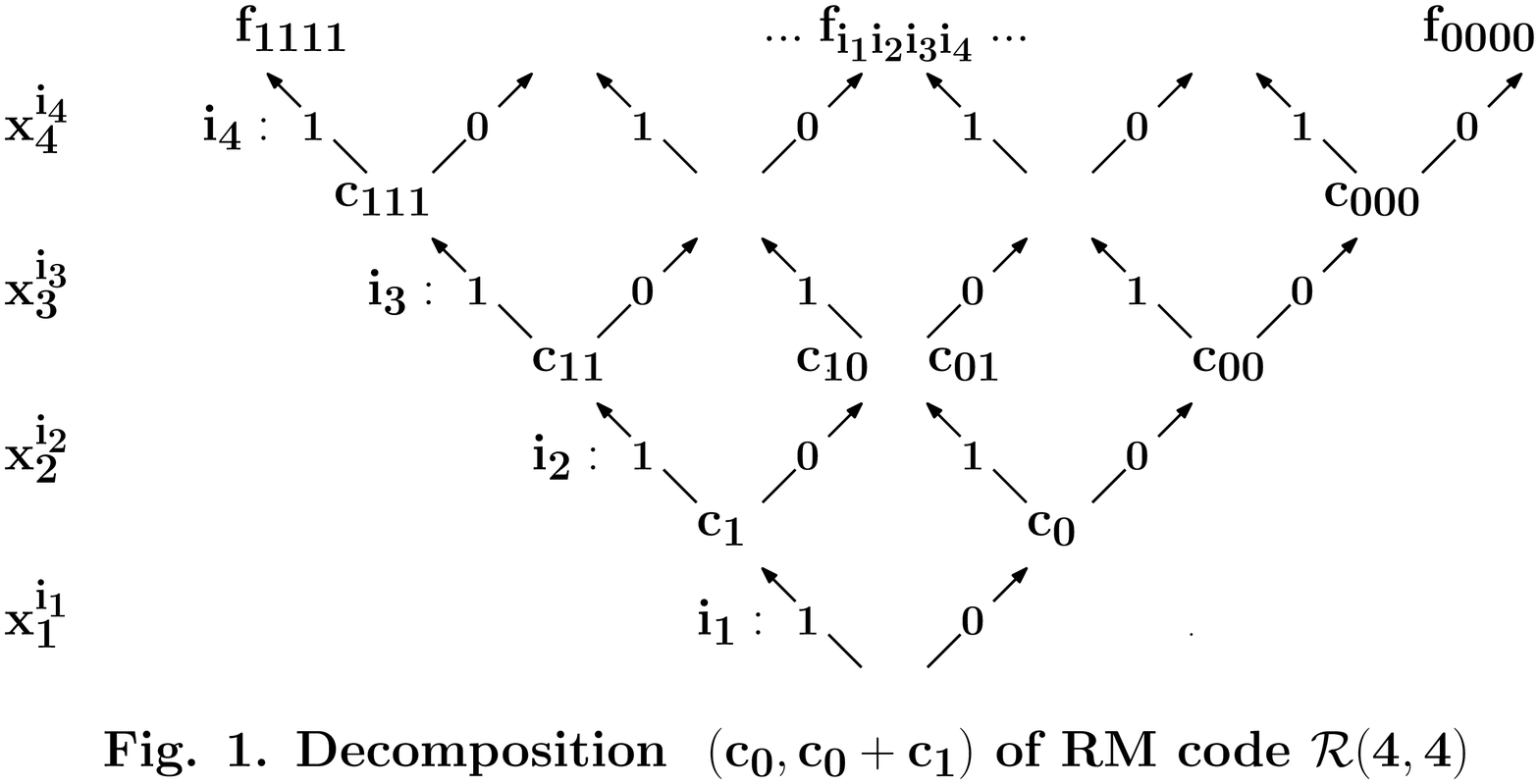}\end{figure}\begin{figure}[ptb]
\hspace{0.5in}%
\includegraphics[width=0.4\textwidth]{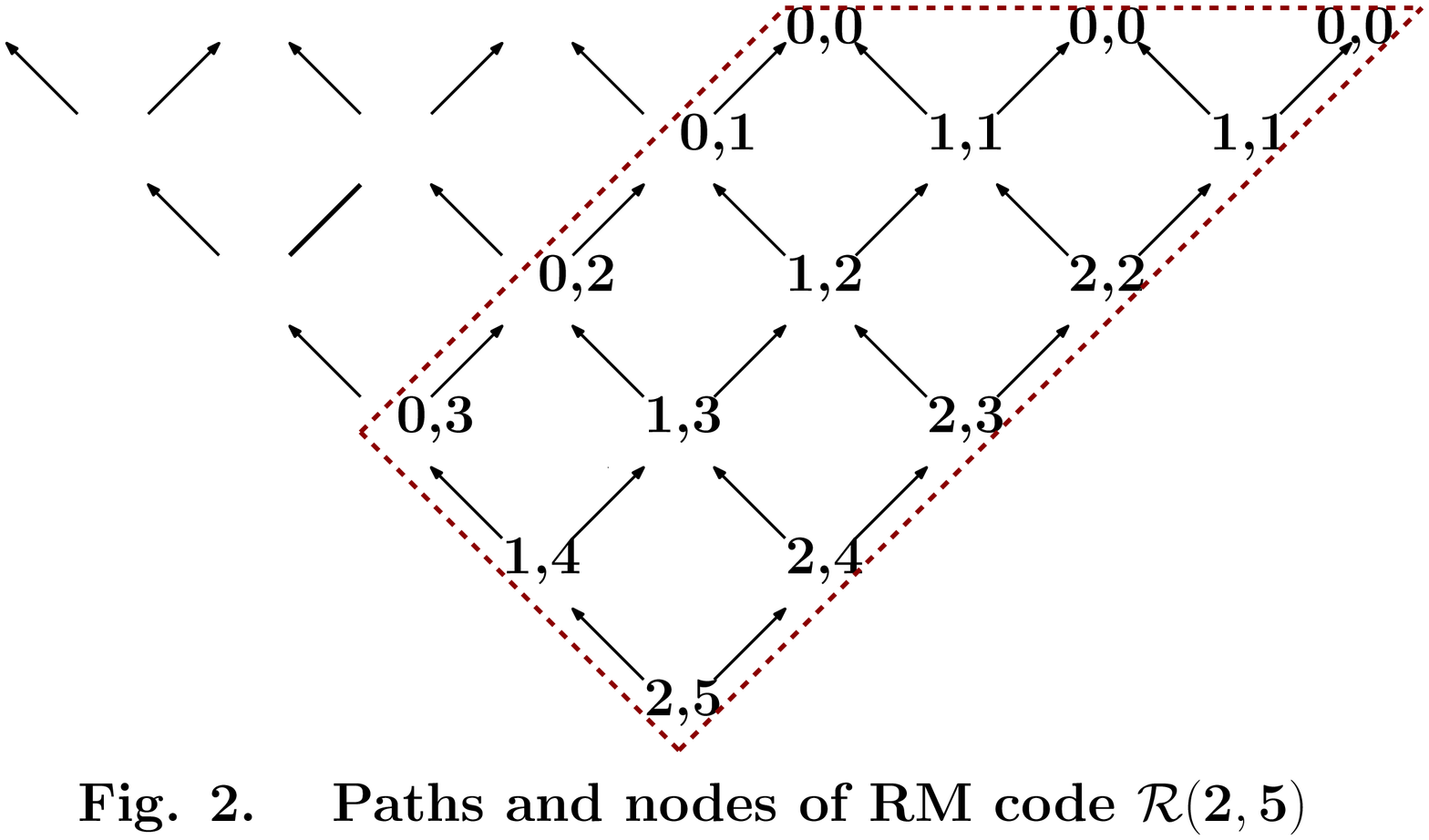}\end{figure}Now consider
some subset of $N$ paths $T.\mathbf{\ }$ Then we encode $N$ information bits
via their paths and obtain codewords $\mathbf{c}(T)=\sum_{\xi\in
T}\mathbf{c\mathbf{(\xi).}}$ These codewords form a linear code $C(m,T).$ Note
also that at any level $\ell$ and at any node $\xi_{\ell\,|\,m}$, encoding
only needs to add two codewords of level $\ell+1$ entering this node$.$ Thus,
encoding performs the order of $2^{m-\ell}$ operations on each of $2^{\ell}$ nodes
$\xi_{\ell\,|\,m}$ and has the overall complexity of $n\log_{2}n$ summed up
over all levels $\ell=1,...,m.$\vspace{0.03in}

\section{\textbf{Recursive decoding algorithms\label{sub:3} }}

\textit{Recursive decoding of RM codes. }Consider a discrete memoryless
channel (DMC) $W$ with inputs $\pm1$ defined by the map $x\rightarrow(-1)^{x}$
for $x=0,1.$ Then we define the codewords $\mathbf{c}=\left(  \mathbf{u,uv}%
\right)  $ of a code $RM(r,m),$ where vector $\mathbf{uv}$ is the
component-wise product of vectors $\mathbf{u}$ and $\mathbf{v}$ with symbols
$\pm1\mathbf{.}$ For any codeword $\mathbf{c},$ let $\mathbf{y}_{0}%
$,$\mathbf{y}_{1}\in\mathbb{R}^{n/2}$ be the two output halves corrupted by
noise. We use double index $i,j$ for any position $j=1,...,n/2$ in a half
$i=0,1$. \ Define the posterior probability (PP) $q_{i,j}=\Pr\{c_{i,j}%
=1\,\,|\,\,y_{i,j}\}$ that $1$ is sent in a position $i,j.$ We will also use
two related quantities, which we call \ \textquotedblleft the
offsets\textquotedblright\ $g_{i,j}$ and the likelihoods $h_{i,j}:$%
\begin{equation}
g_{i,j}=2q_{i,j}-1,\;h_{i,j}=q_{i,j}/\left(  1-q_{i,j}\right)  \label{lik}%
\end{equation}
Thus, we form vectors $\mathbf{q}=(q_{i,j}),$\ \ $\mathbf{g}=(g_{i,j})$ and
$\mathbf{h}=(h_{i,j}).$ The following recursive algorithm $\Psi_{r}%
^{m}(\mathbf{q})$ of \cite{dum1999}, \cite{dum2000} performs SCD of
information bits in codes $RM(r,m)$ or their bit-frozen subcodes $C(m,T).$ We
first derive PP of symbols $v_{j}$ in the $\left(  \mathbf{u,uv}\right)  $
construction:%
\[
q_{j}^{(1)}\equiv\Pr\{v_{j}=1\,\,|\,\,q_{0,j},\,q_{1,j}\}
\]
Namely, it is easy to verify that the offsets $g_{j}^{(1)}$ of symbols
$q_{j}^{(1)}$ satisfy simple recalculations
\begin{equation}
g_{j}^{(1)}=g_{0j}g_{1j}\label{1}%
\end{equation}
We may then apply some decoding algorithm $\Psi_{\,r-1}^{m-1}$ to the vector
$\mathbf{q}^{(1)}\equiv(q_{j}^{(1)})$ and obtain a vector $\widetilde
{\mathbf{v}}\in RM(r-1,m-1)$ of length $n/2.$ Then each half of the vector
($\mathbf{y}_{0},\mathbf{y}_{1}\widetilde{\mathbf{v}})$ forms a corrupted
version of vector $\mathbf{u}$ in the $\left(  \mathbf{u,uv}\right)  $
construction. As a result, every symbol $u_{j}$ of vector $\mathbf{u}$ has the
likelihoods $h_{0,j}$ and $\left(  h_{1,j}\right)  ^{\widetilde{v}_{j}}$ on
these halves. This gives the overall likelihood of every symbol $u_{j}:$%
\begin{equation}
h_{j}^{(0)}=h_{0j}\left(  h_{1j}\right)  ^{\widetilde{v}_{j}}\label{2}%
\end{equation}
Given the vectors $\mathbf{h}^{(0)}\equiv(h_{j}^{(0)})$ and $\mathbf{q}%
^{(0)},$ we can now apply some algorithm $\Psi_{\,r}^{m-1}$ and decode
$\mathbf{q}^{(0)}$ into a vector $\widetilde{\mathbf{u}}$ $\mathbf{\in}$
$RM(r,m-1).$ With respect to polar codes, observe that recalculations
(\ref{1}) degrade the original channel, whereas recalculations (\ref{2})
upgrade it.

Recalculations (\ref{1}) and (\ref{2}) form the level $\ell=1$ of \ SC decoding.
\ We can also use recalculations (\ref{1}), (\ref{2}) for vectors
$\mathbf{q}^{(1)}$ and $\mathbf{q}^{(0)}$ instead of decoding them. Then
levels $\ell=2,...,m$ are processed similarly, moving decoding back and forth
along the paths of Fig. 1 or Fig. 2. For any current path $\eta
=\mathbf{\mathbf{\xi}}_{1\,|\,\ell}$, decoder has an input vector
$\mathbf{q}(\eta)$ that consists of $2^{m-\ell}$ PP. In essence,
this vector represents the output channel of this path $\eta$. Then we process
the $\mathbf{v}$-extension $(\eta,1)$ using recalculations (\ref{1}). After
processing, the node $(\eta,1)$ returns its current output $\widetilde
{\mathbf{v}}(\eta)$ to the node $\eta$. Similarly, we then continue with recalculations (\ref{2}) for the $\mathbf{u}%
$-extension $(\eta,0)$. Thus, $\mathbf{v}$-extensions (marked with 1 on Fig.
1) precede the $\mathbf{u}$-extensions, and all paths $\mathbf{\mathbf{\xi}%
}_{1\,|\,\ell}$ are ordered lexicographically in each step.

Next, consider all full paths $\xi(S)\mathbf{\mathbf{,}}$ $S=1,...,k(r,m).$
Every path $\xi(S)$ ends with one information bit $f\left(  S\right)
\equiv f\left(  \xi(S)\right)$ and gives its likelihood $q\left(  S\right)  =\Pr\{f\left(
S\right)  =0\,\,|\,\,\mathbf{q}\}$.  We then choose the more
reliable value for $f\left(  S\right)  .$ The result is the current sequence
$F(S)=f\left(  1\right)  ,...,f\left(  S\right)  $ of the first $S$
information bits. The decoding ends if $S=k(r,m).$

It is easy to verify \cite{dum1999} that $m$ decomposition steps give
complexity $2n\log_{2}n.$ Indeed, any level $\ell=1,...,m$ includes at most
$2^{\ell}$ paths $\eta=\mathbf{\mathbf{\xi}}_{1\,|\,\ell}.$ Each path $\eta$
\ recalculates vectors $\mathbf{g}(\eta\mathbf{)}$ and $\mathbf{h}(\eta)$ of
length $2^{m-\ell}.$ Recalculations (\ref{1}), (\ref{2}) on these vectors have
complexity order of $2^{m-\ell}.$ Thus, each level $\ell$ of recursion requires the
order of $2^{m}$ operations.\vspace{0.03in}

\textit{Recursive decoding of polar codes. } Any subcode $C(m,T)$ with $N$
ordered paths $\mathbf{\mathbf{\xi}}(1),...,\mathbf{\mathbf{\xi}}(N)$ is
decoded similarly. Here we simply drop all frozen paths $\mathbf{\mathbf{\xi
\notin}}T$ that give information bits $f\left(\xi\right)  \equiv0.$ \ This
gives the following algorithm.\smallskip\smallskip

$\frame{$%
\begin{array}
[c]{l}%
\text{Algorithm }\Psi(m,T)\text{ for code }C(m,T).\smallskip\\
\text{Given: a vector }\mathbf{q}=(q_{i,j})\text{ of PP.}\smallskip\\
\text{Take }S=1,...,N\text{ and }\ell=1,...,m.\smallskip\\
\text{For a path }\mathbf{\mathbf{\xi}}(S)=i_{1}(S),...,i_{m}(S)\text{ in step
}\ell\text{ do:}\smallskip\\%
\begin{array}
[c]{l}%
\text{Apply recalculations}\;\text{(\ref{1}) if }i_{\ell}(S)=1\smallskip
\smallskip\\
\text{Apply recalculations}\;\text{(\ref{2}) if }i_{\ell}(S)=0.\smallskip\\
\text{Output the bit }f\left(  S\right)  \text{ for }\ell=m.
\end{array}
\end{array}
$}\smallskip\smallskip\smallskip$

\textit{Recursive list decoding. }For the bit-frozen subcodes $C(m,T)$ of RM
codes, \textit{list decoding} version $\Psi(m,T,L)$ of this algorithm was
employed in \cite{dum2000}-\cite{dum2006}. Consider processing of any path
$\mathbf{\mathbf{\xi}}(S).$ Then the algorithm already has the list of $L$
most probable code candidates $t=1,...,L$ obtained on the previous paths. Each
candidate is defined by the sequence $F_{t}(S-1)=\left[  f_{t}\left(
1\right)  ,...,f_{t}\left(  S-1\right)  \right]  $ of $S-1$ information bits
and by the current vector
$\mathbf{q}_{t}$ of posterior probabilities derived by processing these $S-1$  paths. 
For each candidate $t$, we then recalculate the vector
$\mathbf{q}_{t}$ on the path $\mathbf{\mathbf{\xi}}(S).$  This is similar to the case $L=1.$ Namely, any intermediate
node $\eta=\mathbf{\mathbf{\xi}}_{1\,|\,\ell}(S)$ of the path
$\mathbf{\mathbf{\xi}}(S)$ is given $L$ most probable vectors $\mathbf{q}%
_{t}(\eta)$ of length $2^{m-\ell}.$ If the path $\mathbf{\mathbf{\xi}}(S)$ has
the new bit $i_{\ell}(S)=1,$ then we follow $\mathbf{v}$-extension $(\eta,1)$
and perform recalculations (\ref{1}) for each vector $\mathbf{q}_{t}(\eta).$
Otherwise, path $\eta$ receives $L$ vectors $\widetilde{\mathbf{v}}_{t}(\eta$)
from the $\mathbf{v}$-extension $(\eta,1)$ and proceeds with recalculations
(\ref{2}) on its $\mathbf{u}$-extension $\left(  \eta,0\right)  .$

Our recalculations are slightly different in the final step $\ell=m.$ Given
the prefix $\eta=\mathbf{\mathbf{\xi}}_{1\,|\,m-1}(S)$ of the path
$\mathbf{\mathbf{\xi}}(S)$, we continue with the same recalculations (\ref{1})
or (\ref{2}) depending on the new symbol $i_{m}(S).$ However, now we consider
both values $f_{t}(S)=0,1$ of a new information bit $f_{t}(S)$. As
a result, we obtain two posterior probabilities $\mathbf{q}_{t}^{\prime
}\mathbf{(}\eta)$ and $\mathbf{q}_{t}^{\prime\prime}\mathbf{(}\eta)$ for each
candidate $t=1,...,L$ on the full path $\mathbf{\mathbf{\xi}}(S).$ Thus, $L$
candidates yield two presorted lists $\{\mathbf{q}_{t}^{\prime}\mathbf{(}%
\eta)\}$ and $\{\mathbf{q}_{t}^{\prime\prime}\mathbf{(}\eta)\}$. To proceed
further, we select $L$ most probable codewords in the combined list, which
requires the order of $\mathcal{O(}L)$ operations.

The result is the new list of information bits $F_{t}(S).$ Note that this list
can exclude some candidates $t$ but keep both values $f_{t}(S)=0,1$ for some
other $t$ until we select the single most probable codeword in the end.  In processing, we also keep the current posterior probabilities
$\mathbf{q}_{t}\mathbf{(\mathbf{\xi}}(S)),$ which will be used in the next
steps for the path $\mathbf{\mathbf{\xi}}^{(S+1)}.$   
A more detailed
description of this decoding algorithm is also given in \cite{dum2006}.

Note that each level $\ell$ includes at most $2^{\ell}$ nodes
$\mathbf{\mathbf{\xi}}_{1\,|\,\ell},$ each of which processes $2L$ vectors of
length $2^{m-\ell}.$ Given some constant number $c$ operations per code
symbol, we only perform $2cLn$ operations in step $\ell,$ and then relegate decoding to
step $\ell+1.$ Thus, we can bound complexity $\left\vert \Psi_{\ell
}(m,T,L)\right\vert $ of level $\ell$ and the overall complexity $\left\vert
\Psi(m,T,L)\right\vert $ of the list decoding by the order of $Ln\log_{2}n$:%
\[
\left\vert \Psi_{\ell}(m,T,L)\right\vert \leq\left\vert \Psi_{\ell
+1}(m,T,L)\right\vert +2cLn
\]%
\[
\left\vert \Psi(m,T,L)\right\vert \leq\sum\nolimits_{\ell=1}^{m}%
2cLn=2cLn\log_{2}n
\]
This concludes our description of the list decoding algorithm.

\end{document}